\documentclass{article}
\usepackage{graphicx}
\usepackage{hyperref}
\usepackage{amsmath}
\usepackage{amssymb}	% Extra maths symbols
\usepackage{mathtools}

\usepackage{subcaption}
\usepackage[dvipsnames]{xcolor}

\usepackage{multirow}
\usepackage{pdfpages}
\usepackage{stfloats}
\usepackage{tabularx}
\usepackage[export]{adjustbox}
\usepackage{float}
\usepackage[
singlelinecheck=false % <-- important
]{caption}

\usepackage{multicol}

\begin{document}
\title{\bf Time-dependent wormhole solutions in conformal Weyl gravity }
\author{{Malihe Heydari-Fard\thanks{Electronic address:\href{mailto:heydarifard@qom.ac.ir}{heydarifard@qom.ac.ir}}, Mohammad Rahim Bordbar\thanks{Electronic address:\href{mailto:mbordbar@qom.ac.ir}{mbordbar@qom.ac.ir}} and Golnaz Mohammadi\thanks{Electronic address:\href{mailto:Golnaz.6.1364@gmail.com}{Golnaz.6.1364@gmail.com}}}\\ {\small \emph{ Department of Physics, The University of Qom, 3716146611, Qom, Iran}}
}
\maketitle

\begin{abstract}
We present the exact time-dependent solutions on inhomogeneous spherically symmetric space-time in the conformal invariant Weyl gravity. For this purpose, the subclass of the Lemaitre-Tolman metric which is supported by an anisotropic fluid is used. For the first time, the exact solutions of the dynamical equations are obtained for two special cases. One of the exact solutions is a de Sitter space-time and other solution is a class of time-dependent wormhole geometries which can be supported by exotic matter in similar to the general relativistic solutions.
\vspace{5mm}
\end{abstract}

\begin{multicols}{2}
\section{Introduction}
\label{introduction}
The first solution for wormholes was studied by Flamm in the context of General Relativity (GR) \cite{L. Flamm} in 1930. Later in 1933 Weyl \cite{H. Weyl} introduced a wormhole as a tunnel-like structure lying in the same universe or linking two remotely separated regions. In the mid-1935, Einstein and Rosen constructed an
unstable wormhole known as the Einstein-Rosen bridge \cite{A. Einstein and N. Rosen}. Morris and Thorne proposed a solution to Einstein's field equation by imposing
inhomogeneous and static spherical symmetry on space-time. Their solutions (called traversable wormholes) were topological objects with
a throat connecting two asymptotically flat regions \cite{Morris:1988tu}. Traversable wormholes have no horizons or curvature singularities but they made
gravitational forces that are assumed to be bearable by travelers. The most important characteristic of these wormholes is the throat,
where the exotic matter should be existed to prevent them from collapsing. Indeed such space-time requires a stress-energy tensor
that violates the null energy condition \cite{Hochberg:1998qw}.

From the theoretical point of view in quantum field theory, the possibility of exotic energy has been accepted in the Casimir effect \cite{Casimir}.
Interesting evidence of the experimental effect is an attractive force between two parallel metallic plates in a vacuum that is generated by
exotic matter \cite{avino2019progress}. Wormholes with negative energy densities in quantum gravity have been studied using the path integral in \cite{hebecker2018euclidean,Mertens}. Moreover, Hochberg and Kephart have indicated that the wormhole geometry with negative energy might be produced by the gravitational squeezing of the vacuum \cite{hochberg1991lorentzian}. Another field in which we deal with exotic matter is cosmology. Based
on \cite{narlikar2002introduction}, the exotic matter with the property of $w < -1/3$ and the equation of state $p = w \rho$ is responsible for accelerating the universe. Phantom energy with $w <-1$ has some other properties such as negative temperature and energy and Big Rip whose energy
density evolves with expanding universe \cite{caldwell2003phantom}. Sato et al. \cite{Sato:1981bf} have investigated the possibility of dynamical
wormhole formation in the inflationary era. Other aspects of evolving wormholes of the Planck length scale have been considered by Friedman \cite{Friedman:1988qs} and Roman \cite{Roman:1992xj}.

Although wormholes are explained by Einstein's gravity, there are still traversable wormhole solutions in alternative theories of gravity. Wormhole solutions have been studied in Brans–Dicke theory \cite{Agnese:1995kd}--\cite{MontelongoGarcia:2011ag}, $f(R)$ \cite{fR}--\cite{R2}, $f(R, T)$ gravity \cite{moraes2017modeling}, massive gravity \cite{massive}, scalar-tensor theory \cite{Accetta:1989gh, Shaikh:2016dpl}, third order Lovelock \cite{KordZangeneh:2015dks} and Kaluza–Klein gravity \cite{Dzhunushaliev:1999aka}. Also static wormholes in the presence of a cosmological constant and Born-Infeld theory have been reported in \cite{Lemos:2003jb} and \cite{MontelongoGarcia:2010xd, Garcia:2010xb}, respectively. In Ref. \cite{moraes2019exponential} authors have studied the wormholes in $f(R, T)$ modified gravity theory by using an exponential shape function. In Einstein-Gauss-Bonnet (EGB) gravity, vacuum wormhole solutions have been obtained \cite{GB1}--\cite{dotti2007static}. Also the higher-dimensional wormholes has been of interest in the last years \cite{paul2021emergent, oliveira2022traversable}. Investigation of classical wormholes based on conformal Weyl gravity has been done by Varieschi and Ault \cite{varieschi2016wormhole}. Some works on the subject already exists in the literature (see for example \cite{F. S. N. Lobo,J. Oliva}). Time-dependent wormhole solutions on inhomogeneous and spherically symmetric space-time in the presence of matter source with radial and transverse stresses have been obtained in \cite{Bordbar:2010zz}. Also by considering an inhomogeneous brane embedded in 5-dimensional constant curvature bulk, time-dependent wormhole solutions as exact solutions on the brane have been found in \cite{Heydari-Fard:2017oee}. For the study of the evolving Lorentzian wormholes and the null energy condition (NEC) an weak energy condition (WEC) see \cite{ev1}--\cite{ev8}. Also evolving wormholes in $f(R)$ gravity theory, Einstein-Cartan gravity, EGB gravity, Lovelock gravity and Rastall theory were investigated in Ref. \cite{fr}--\cite{Rastall}, respectively.

Despite the incredible successes of GR theory, there are basic problems in astronomy and cosmology, such as dark energy and dark matter. An alternative approach to describe the cosmic structure of the universe without considering dark matter is modifying the theory of gravity. Amongst the many modifications of GR, conformal invariant Weyl gravity is proposed in 1918 by Weyl \cite{W1, W2} and developed by Bach \cite{W3}. Although finding exact solution of this fourth-order conformal Weyl gravity is a formidable endeavor, the exact vacuum solution to conformal Weyl gravity and its implications have been studied by Mannheim and Kazanas (MK metric) \cite{Mannheim:1988dj, Kazanas:1988qa}.This exact static spherically symmetric vacuum solution is given by the following metric
$$
ds^2 = -B(r)^2 dt^2+\frac{1}{B(r)}dr^2+r^2(d{\theta}^2+sin^2 \theta d{\varphi}^2 ),
$$
where
$$
B(r) = 1-\frac{\beta(2-3\beta\gamma)}{r}-3\beta\gamma+\gamma r-k r^2,
$$
and $\beta$, $\gamma$ and $k$ are integration constants. This exterior solution includes three new extra terms to the standard Schwarzschild metric which can explain the observed galactic rotation curves without introducing dark matter \cite{Mannheim:1999bu, Edery:2001at}. Cylindrically symmetric solutions in conformal gravity were presented in Ref. \cite{cy2, cy3, cy4}. Dynamical cylindrical symmetric solutions in conformal Weyl gravity have been investigated in \cite{cy1}. The purpose of this paper is to find the dynamical spherically symmetric solutions in the framework of the conformal Weyl gravity.

The exact solutions to the Reissner-Nordstrom, Kerr and Kerr-Newmann space-times have been studied in \cite{Mannheim:1990ya}. An interesting application of fourth-order conformal Weyl gravity is analysis the traversable wormhole solutions in this theory. The wormhole solutions in the theory of GR, are supported by exotic matter which violates main energy conditions \cite{Visser}; so an interesting challenge in wormhole physics is the demand to find a realistic matter that will support these exotic space-times. The computation of light bending angle by a spherically symmetric object using MK metric, has been studied in detail \cite{Edery}--\cite{Pireaux:2004id}. For an asymptotically non-flat geometry such as MK metric by using Rindler-Ishak method, the total light deflection angle to second order has been calculated in \cite{Sultana:2010zz} and \cite{Sultana:2013fda,Bhattacharya:2010xh}. Correct light deflection in Weyl conformal gravity has been obtained in \cite{Cattani:2013dla}. In \cite{Sultana} authors have investigated the perihelion shift of planetary motion in conformal Weyl gravity. For astrophysical tests in conformal Weyl gravity see \cite{Varieschi:2014pca,Dutta}.

The structure of paper is as follows. In section \ref{sec1}, after a brief review of Weyl gravity, the Szekeres-Szafron metric is introduced. Then we obtain the field equations for this inhomogeneous space-time in the framework of conformal Weyl gravity in subsection \ref{sec1-3}. Finally we solve these equations to find two physical and important solutions in section \ref{sec3}. The paper ends with concluding remarks in section \ref{sec4}.

\section{Field equations in Conformal Weyl gravity
\label{sec1}}

\subsection{Weyl action\label{sec1-1}}
Conformal Weyl gravity is based on the following action \cite{DeWitt}
\begin{eqnarray}
I_{w} &=& -\alpha \int d^{4}x \sqrt{-g}  C_{\lambda\mu\nu\kappa} C^{\lambda\mu\nu\kappa} \nonumber\\
&=& -2\alpha  \int d^{4}x \sqrt{-g} \left (R_{\mu\nu} R^{\mu\nu} - \frac{1}{3} R^{2}\right),
\label{1}
\end{eqnarray}
where $g \equiv \det(g_{\mu\nu})$, $\alpha$ is the  coupling
constant and
\begin{eqnarray}
C_{\lambda\mu\nu\kappa} &=& R_{\lambda\mu\nu\kappa}-g_{\lambda[\nu} R_{\kappa]\mu}+ g_{\mu[\nu} R_{\kappa]\lambda} \nonumber\\
&+&\frac{1}{2}R g_{\lambda[\nu}g_{\kappa]\mu},
\label{Weyl}
\end{eqnarray}
is the Weyl tensor \cite{H. Weyl}.

By varying the action (\ref{1}) with respect to the $g_{\mu\nu}$, we obtain the following field equations
\begin{equation}
2\alpha W_{\mu\nu} = \frac{1}{2} T_{\mu\nu},
\label{0}
\end{equation}
or
\begin{equation}
2\alpha\left [ -\frac{1}{3}W_{\mu\nu}^{(1)}+W_{\mu\nu}^{(2)}\right] = \frac{1}{2} T_{\mu\nu},
\label{2}
\end{equation}
where
\begin{equation}
W_{\mu\nu} \equiv W_{\mu\nu}^{(2)} -\frac{1}{3}W_{\mu\nu}^{(1)},
\label{n}
\end{equation}
here $W_{\mu\nu}^{(1)}$ and $W_{\mu\nu}^{(2)}$ are defined as
\begin{equation}
W_{\mu\nu}^{(1)} = 2 g_{\mu\nu} R^{;\beta}_{\,\,\,\,;\beta}-2 R_{;\mu \nu} -2 R R_{\mu\nu}+ \frac{1}{2}g_{\mu\nu} R^{2},
 \label{3}
\end{equation}
and
\begin{eqnarray}
 W_{\mu\nu}^{(2)} &=& \frac{1}{2} g_{\mu\nu}R^{;\alpha}_{\,\,\,\,;\alpha}+R^{\,\,\,\,;\beta}_{\mu\nu\,\,\,\,;\beta}-R_{\mu ;\nu \beta}^{\,\,\,\,\beta}-
 R^{\,\,\,\,\beta}_{\nu ;\mu \beta}\nonumber\\
 &-&2 R_{\mu\beta}R^{\beta}_{\,\,\,\,\nu}+\frac{1}{2}g_{\mu\nu}R_{\alpha\beta}R^{\alpha\beta},
\label{4}
\end{eqnarray}
respectively.

In equation (\ref{0}) $ W_{\mu \nu}$ and $T_{\mu \nu}$ are symmetric, traceless and covariantly conserved. We will use these properties in the next sections. Also the energy-momentum tensor is defined as
\begin{equation}
T_{\mu \nu} = -\frac{2}{\sqrt{-g}}\frac{\delta(\sqrt{-g} L_m)}{\delta (g^{\mu\nu})},
\label{5}
\end{equation}
where $L_{m}$ is the matter Lagrangian density.

\subsection{Space-time geometry\label{sec1-2}}

In commoving coordinate, the form of Szekeres-Szafron metric is \cite{60}--\cite{62}
\begin{equation}
ds^2  = -dt^2  + R(t)^2 \left[(1 + a(r))dr^2+ r^2 d{\theta}^2+ r^2sin^2 \theta d{\varphi}^2 \right],
\label{7}
\end{equation}
here $R(t)$ is the cosmic scale factor and $a(r)$ is an unknown function of the radial coordinates of $r$. Note that we use our metric signature is $(-,+,+,+)$. We have the Robertson-Walker (RW) metric as a special case
\begin{equation}
1+a(r)=\frac{1}{1-kr^2},
\label{8}
\end{equation}
where $k$ is the spatial curvature index which take the values:$-1,0,1;$ corresponding to the open, flat and closed cases, respectively.

\subsection{Field equations\label{sec1-3}}
Now by inserting the metric (\ref{7}) in the field equations (\ref{0}), we obtain
\begin{equation}
W_{t}^{t}=\frac{f(r)}{R(t)^4},
\label{12}
\end{equation}

\begin{equation}
W_{r}^{r}=\frac{g(r)}{R(t)^4},
\label{13}
\end{equation}

\begin{equation}
W_{\theta}^{\theta}=W_{\varphi}^{\varphi}=\frac{h(r)r^2}{R(t)^4},
\label{14}
\end{equation}
where $f(r)$, $g(r)$ and $h(r)$ are defined as
\begin{eqnarray}
f(r)&=&\frac{1}{12{(a+1)^5 r^4}} [(4a^2 a^{\prime\prime\prime}- 26 a  a^{\prime} a^{\prime\prime}\nonumber\\
&+&28 a^{\prime 3}-26 a^{\prime} a^{\prime\prime}+4 a^{\prime\prime\prime}+8a a^{\prime\prime\prime}) r^3 \nonumber\\
&+&(4a^2  a^{\prime\prime}-7aa^{\prime 2}+ 8a a^{\prime\prime}- 7 a^{\prime 2}\nonumber\\
&+&4 a^{\prime\prime}) r^2-(8a^2  a^{\prime}+16 a a^{\prime}+8 a^{\prime}) r  \nonumber\\
&&+20a^4+36a^3+28a^2+8a],
\label{9}
\end{eqnarray}
\begin{eqnarray}
g(r)&=&\frac{1}{12 (a+1)^4 r^4}[(-4aa^{\prime\prime}+7 a^{\prime 2}-4 a^{\prime\prime})r^2 +4a^4\nonumber\\
&+& 16a^3 +20a^2 +8a],
\label{10}
\end{eqnarray}
\begin{eqnarray}
h(r)&=& -\frac{1}{12{(a+1)^5 r^6}
}[(2a^2 a^{\prime\prime\prime}-13 a a^{\prime} a^{\prime\prime}\nonumber\\
&+&14a^{\prime 3}+4 a a^{\prime\prime\prime}-13 a^{\prime}a^{\prime\prime}+2a^{\prime\prime\prime}) r^3\nonumber\\
&+&(-4a^2 a^{\prime}-8a a^{\prime}-4 a^{\prime}) r+ 4 a^5\nonumber\\
&+&20a^4+ 36a^3+28 a^2 +8a],
\label{11}
\end{eqnarray}
where the prime denotes to the derivative with respect to the radial coordinate $r$.

The energy-momentum tensor required to support such a space-time is in the form,
\begin{equation}
T^{\mu}_{\nu} = \rm diag(-\rho ,P_{r},P_{t}, P_{t}),
\label{15}
\end{equation}
where $\rho(r,t)$ is the energy density and  $P_{r}(r,t)$, $P_{t}(r,t)$ are the radial and transverse pressures, respectively. Use of equations (\ref{12})-(\ref{14})
and equation (\ref{15}) and substituting into equation (\ref{0}) lead to the following equations
\begin{equation}
\rho(r,t) = -4\alpha \frac{f(r)}{R(t)^4},
\label{16}
\end{equation}
\begin{equation}
P_{r}(r,t) = 4\alpha\frac{g(r)}{R(t)^4},
\label{17}
\end{equation}
\begin{equation}
P_{t}(r,t) = 4\alpha\frac{h(r)r^2}{R(t)^4}.
\label{18}
\end{equation}
To calculate $a(r)$ and $R(t)$ in Weyl gravity, we use two properties of Weyl's tensor, Bianchi and trace identities as
\begin{equation}
\nabla_{\mu} W^{\mu\nu} =0,
\label{19}
\end{equation}
\begin{equation}
 W^{\mu}_{\mu}=0,
 \label{20}
\end{equation}
use of equation (\ref{19}) for $\nu = t$ leads to
\begin{equation}
\frac{1}{R(t)} \frac{d{R}(t)}{dt}\left[\frac{f(r)+g(r)+2h(r){r^2}}{R(t)^4}\right] =0,
\label{21}
\end{equation}
and for $\nu=r$ from equation (\ref{19}), we have
\begin{equation}
g^{\prime}(r)+\frac{2g(r)}{r}-2rh(r)=0.
\label{22}
\end{equation}
Also, we use the trace identity (\ref{20}) to obtain
\begin{equation}
\frac{f(r)+g(r)+2h(r){r^2}}{R(t)^4}=0,
\label{23}
\end{equation}
or
\begin{equation}
-\rho+P_{r}+2 P_{t}= 0.
\label{con}
\end{equation}
As we know there are two unknown functions  $R(t)$ and  $a(r)$ to obtain the metric, as well as  $\rho$,  $P_{t}$ and  $P_{r}$, are unknown and functions of $r$ and $t$. In the standard GR \cite{Bordbar:2010zz} and the brane-world model \cite{ Heydari-Fard:2017oee} in order to obtain the inhomogeneous exact solutions, authors have chosen the generalized equation of state as follow
\begin{equation}
\rho+\alpha P_{r}+2 \beta P_{t}=0.
\label{24}
\end{equation}
where $\alpha$ and $\beta$ are constant parameters. But, we note that in the conformal Weyl gravity the energy-momentum tensor components are constrained through the trace identity
(\ref{con}), which means $\alpha = \beta = -1$. Thus in contrast to standard GR, we can not use equation (\ref{24}) to obtain a new equation to find $a(r)$ and $R(t)$.

Weyl equations (\ref{16})-(\ref{18}) together with equations (\ref{21}), (\ref{22}), and (\ref{23}) make a set of equations which can be solved. In the next section by imposing constrain between the radial and transverse pressures, we obtain exact solutions for $R(t)$ and $a(r)$.

\section{Exact solutions in Weyl gravity\label{sec3}}
In this section we are going to obtain inhomogeneous exact solutions in the framework of Weyl gravity. As we know the equation of state has an important role in the study of the geometry of space-time. For example $\omega=-1$ correspond with the vacuum energy or cosmological constant and $-1<\omega<-1/3$ are  mentioned for the quintessence matter and used as a candidate for explaining the accelerated expansion of the universe. Phantom field as an exotic matter with equation of state parameter $\omega<-1$ also accelerate the expansion of the universe.

\subsection{case I: Isotropic fluid}
First we focus on the cosmic scale factor $R(t)$. By comparing two equations (\ref{21}) and (\ref{23}), we conclude that
\begin{equation}
\frac{\dot{R}(t) }{R(t)}\neq 0.
\label{26}
\end{equation}
The above equation shows that there are different choices to get the scale factor. So the scale factor can be an arbitrary function of time. Inflating Lorentzian wormholes in the framework of GR were investigated by Roman \cite{Roman:1992xj} which explore the possibility that inflation provide a natural mechanism for the enlargement wormholes from microscopic size to macroscopic. For having an exponential inflation we consider the simplest choice. By choosing $\frac{\dot{R}(t)}{R(t)} = \rm constant$, we have the following solution for $R(t)$
\begin{equation}
R(t) = R_{0} e^{H_{0}t},
\label{27}
\end{equation}
where $H_{0}$ is the constant of integration.

Now, we consider an isotropic fluid
\begin{equation}
P_{r}(r,t) = P_{t}(r,t),
\label{28}
\end{equation}
 which gives
\begin{equation}
g(r) = h(r)r^2.
\label{28}
\end{equation}
Substituting equation (\ref{28}) into equation (\ref{22}) we obtain
\begin{equation}
r g^{\prime}(r) = 0,
\label{29}
\end{equation}
which have the following solution
\begin{equation}
g(r) = c_1,
\label{30}
\end{equation}
where $c_1$ is an integration constant.

By substituting $g(r)$ from equation (\ref{30}) into equation (\ref{10}), we obtain
\begin{eqnarray}
12 c_1 {(a+1)^4 r^4} &=& 4a^4-4a a^{\prime\prime}r^2+7 a^{\prime 2} r^2+16a^3
\nonumber\\
&-&4 a^{\prime\prime}r^2+20a^2+8a.
\label{31}
\end{eqnarray}
The above equation has not an exact solation. For the case $c_1 = 0$ we find the following exact solution as
\begin{equation}
a(r)=\frac{c_2 r^2}{1 - c_2 r^2} ,
\label{32}
\end{equation}
that $c_2$ is a constant of integration. Now from equations (\ref{27}) and (\ref{32}), the line element (\ref{7}) takes the form
\begin{eqnarray}
ds^2=-dt^2+R_{0}^2 e^{2 H_0 t}\left[\frac{dr^2}{1-c_2 r^2}+r^2 d{\Omega}^2\right],
\label{33}
\end{eqnarray}
where $d{\Omega}^2 = d\theta^2+\sin^2\theta d\varphi^2$. For this case from equations (\ref{16})-(\ref{18}) we find $\rho = P_r = P_t = 0$, which is the simplest case which satisfying the trace equation (\ref{con}).

The spatial part of metric (\ref{33}) shows an exponentially expanding $3$-sphere, and therefore describes a closed empty universe for $c_2 > 0$ and a open empty universe for $c_2 < 0$.

Also for the special case $c_2 = 0$ it corresponds  to
\begin{equation}
ds^2 = -dt^2+R_{0}^2 e^{2 H_0 t}\left[{dr^2}+r^2 \left(d\theta^2+\sin^2\theta d\varphi^2\right)\right],
\label{333}
\end{equation}
which presents the de Sitter space-time.

Therefore as mentioned in Ref. \cite{cy1} the conformal invariance imposes so sharp constraint on isotropic distributions of matter in the universe; so that in an empty FRW universe, the scale factor can be an arbitrary function of time. The merit of this is that we don't need any exotic matter to explain the acceleration expansion of the universe \cite{cy1}.

\subsection{case II: Anisotropic fluid}
Now, we consider the case when the following relation between the energy density $\rho(r)$ and the radial pressure $P_{r}(r)$
\begin{equation}
P_{r} (r,t) = \omega \rho(r,t),
\label{n1}
\end{equation}
where $\omega$ is the equation of state parameter.

Substituting equations (\ref{16}) and (\ref{17}) into equation (\ref{n1}), we have
\begin{equation}
g(r) = -\omega f(r),
\label{3444}
\end{equation}
by omitting $h(r)$ between equation (\ref{22}) and equation (\ref{23}), we have
\begin{equation}
r g^{\prime}(r) +3 g(r)= -f(r),
\label{344}
\end{equation}
by combining equations (\ref{3444}) and (\ref{344}), we have
\begin{equation}
r g^{\prime}(r) = \frac{(1-3\omega)}{\omega} g(r),
\label{n2}
\end{equation}
with the following solution
\begin{equation}
g(r) = c_1 r^{\frac{(1-3\omega)}{\omega}},
\label{n3}
\end{equation}
that $c_1$ is an integration constant. By substituting $g(r)$ from equation (\ref{n3}) into equation (\ref{10}), we obtain
\begin{eqnarray}
12 c_1 r^{\frac{(1-3\omega)}{\omega}} {(a+1)^4 r^4} &=& 4a^4-4a a^{\prime\prime}r^2+7 a^{\prime 2} r^2+16a^3\nonumber\\
&-&4 a^{\prime\prime}r^2+20a^2+8a.
\label{34}
\end{eqnarray}
In general the above equation could not be solved unless we set $\omega =-1$. Unfortunately, even in this case, the equation does not have an explicit form of the exact solution. However, in Appendix A, we present a solution containing an integration term with three constants of integration $c_1$, $c_2$ and $c_3$. For different values of these constants there are many different solutions, however, some of them don't have the physical meaning. Now, in what follows we consider the case of $c_1 = \frac{1}{3}$ which leads to a solution satisfying all of the wormhole conditions.

\subsubsection{The case of $c_1 = \frac{1}{3}$}
As is clear from equation (\ref{Ap1}), by choosing $c_1 = \frac{1}{3}$ the integrand takes a simple form and thus one can easily find the following exact solution
for $a(r)$ function. Also, by choosing $c_1 = \frac{1}{3}$ in equation (\ref{34}) we have
\begin{eqnarray}
4 a a^{\prime\prime}r^2+4 a^{\prime\prime}r^2-7 a^{\prime 2} r^2+8a+4a^2+4 = 0.
\label{35}
\end{eqnarray}
It can be shown that this equation has the following exact solution
\begin{equation}
a(r)=-1+\frac{1}{(\frac{3}{8})^{\frac{4}{3}}\left(c_2 r^{\frac{3}{2}}-c_3 r^{-\frac{1}{2}}\right)^{\frac{4}{3}}},
\label{36}
\end{equation}
where $c_2$ and $c_3$ are integration constants. The line element (\ref{7}) takes the form
\begin{equation}
ds^2=-dt^2+R_{0}^2e^{\sqrt{\frac{\Lambda}{3}}t}\left[\frac{dr^2}{(\frac{3}{8})^{\frac{4}{3}}\left(c_2 r^{\frac{3}{2}}-c_3 r^{-\frac{1}{2}}\right)^{\frac{4}{3}}}+r^2 d{\Omega}^2\right],
\label{37}
\end{equation}
where $H_0 \equiv {\sqrt{\frac{\Lambda}{3}}}$.

The time-dependent wormholes have been introduced by Roman with the following line-element \cite{Roman:1992xj}
\begin{equation}
ds^2=-dt^2+R^2(t)\left[\frac{dr^2}{1-\frac{b(r)}{r}}+r^2 \left(d\theta^2+\sin^2\theta d\varphi^2\right)\right],
\label{38}
\end{equation}
where $R(t)$ and $b(r)$ are the scale factor and the shape function of wormhole, respectively \cite{Morris}. The minimum value of $r$ is a throat radius of wormhole $r=r_0$, so
the radial coordinate change in the interval $r_0 \leq r \leq\infty $. Since the shape function $b(r)$ is responsible to define the shape of the
wormhole, hence for a wormhole solution it should satisfy the certain conditions: i) The radius of the wormhole throat corresponds with the point where $b(r_0) = r_0$, ii) The flaring-out condition implies that $b^{\prime}(r)<1$ and iii) For $r>r_0$ the throat condition imply that $\frac{b(r)}{r} <1$ (for more study the reader is referred to \cite{Lemos}--\cite{Dehghani:2009zza}).

\begin{figure}[H]
\centering
\includegraphics[width=3in]{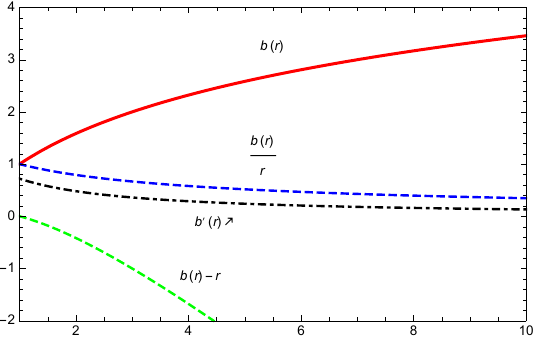}
\caption{Shape function $b(r)$, throat condition $\frac{b(r)}{r} < 1$, flaring-out condition $b^{{\prime}}(r) < 1$ for throat radius $r_{0} = 1$ , $c_2 = 0.1$ and $c_3 = 0.1$.}
\label{br}
\end{figure}

Comparison of metric (\ref{37}) with (\ref{38}) leads to the following shape function
\begin{equation}
b(r) = r-{\left(\frac{3}{8}\right)^{\frac{4}{3}}\left(c_2 r^{\frac{9}{4}}-c_3 r^{\frac{1}{4}}\right)^{\frac{4}{3}}},
\label{35}
\end{equation}
and from condition (i) the throat radius is
\begin{equation}
r_0 = \left(\frac{c_3}{c_2}\right)^{\frac{1}{2}},
\label{39}
\end{equation}
which is real only if ($c_2 > 0$, $c_3 > 0$). One can find $c_3$ in terms of $r_0$, $c_2$ ; so we rewrite the shape function as
\begin{equation}
b(r) = r-{\left(\frac{3}{8}\right)^{\frac{4}{3}}\left(c_2  r^{\frac{9}{4}}-c_2 r_0^2 r^{\frac{1}{4}}\right)^{\frac{4}{3}}},
\label{40}
\end{equation}
in this case. In Fig.\ref{br} we have plotted the shape function with various conditions. As the figure shows all necessary conditions are satisfied by the given shape function.

\begin{figure}[H]
\centering
\includegraphics[width=3in]{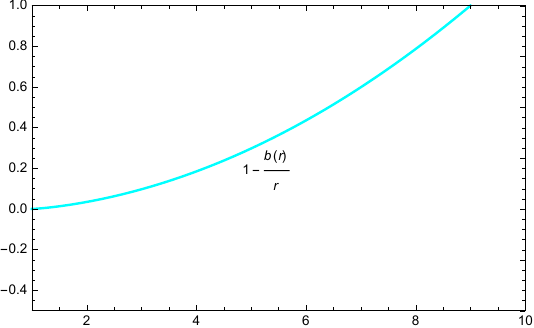}
\caption{ The behavior of $1-\frac{b(r)}{r}$ as a function of $r$.}
\label{fR}
\end{figure}

Quasi-cosmological traversable wormhole solutions in the context of $f(R)$ gravity have been studied in Ref.\cite{fR}. In contrast to the GR one can find the asymptotically spherical, flat and hyperbolic wormhole solutions in modified gravity theories. We have plotted the behavior of function $(1-\frac{b(r)}{r})$ in Fig.\ref{fR}. It shows that the wormhole solutions in Weyl gravity at large $r$ match the hyperbolic FRW universe and so the asymptotically flatness condition is violated.

As we mentioned before the traversable wormholes violate the some main energy conditions such as NEC, WEC, strong energy condition (SEC) and dominated energy condition (DEC) for the stress-energy tensor and so they invoke the existence of exotic matter i.e., matter with negative energy density places at or near the wormhole throat. However, in higher dimensional theories, $f(R)$  gravity theories and modified gravity theories with higher order curvature terms, the wormhole solutions may satisfy some energy conditions \cite{fR}--\cite{R2} and \cite{Parsaei}.

By substituting equations (\ref{36}) into equations (\ref{9})–(\ref{11}), we have
\begin{equation}
f(r) = +\frac{1}{3 r^4},
\label{f43}
\end{equation}
\begin{equation}
g(r) = +\frac{1}{3 r^4},
\label{g44}
\end{equation}
\begin{equation}
h(r) = -\frac{1}{3 r^4},
\label{h45}
\end{equation}
Now we can obtain the energy density, the radial and transverse pressure by substituting equations (\ref{f43})–(\ref{h45}) and equation (\ref{27}) into equations (\ref{16})–(\ref{18}) as follows
\begin{equation}
\rho(t,r) = -\frac{4 \alpha}{3R_0^4 }\frac{1}{r^4} {e^{-\sqrt{\frac{16\Lambda}{3}}t}},
\label{43}
\end{equation}
\begin{equation}
P_{r}(t,r) = +\frac{4 \alpha}{3R_0^4 }\frac{1}{r^4}{e^{-\sqrt{\frac{16\Lambda}{3}}t}},
\label{44}
\end{equation}
\begin{equation}
P_{t}(t,r) = -\frac{4 \alpha}{3R_0^4 }\frac{1}{r^4}{e^{-\sqrt{\frac{16\Lambda}{3}}t}},
\label{45}
\end{equation}
where $H_0 \equiv {\sqrt{\frac{\Lambda}{3}}}$.

\begin{figure}[H]
\centering
\includegraphics[width=3in]{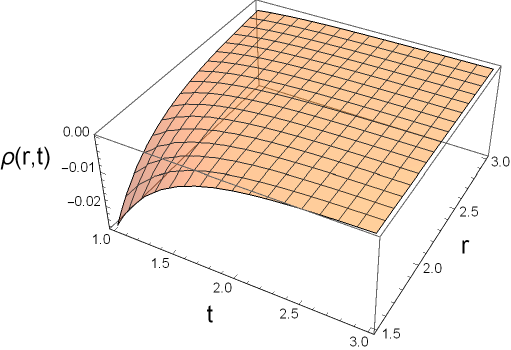}
\caption{The variation of WECs ($\rho(r,t)$) for $R_0 = \alpha = \Lambda = 1$ and the throat radius $r_0 = 1$.}
\label{EC}
\end{figure}

\begin{figure}[H]
\centering
\includegraphics[width=3in]{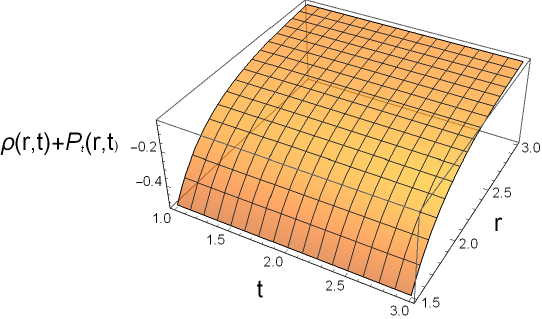}
\caption{The variation of NECs ($\rho(r,t)+P_{t}(r,t)$) for $R_0 = \alpha = 1$ and $\Lambda = 10^{-35}$ and the throat radius $r_0 = 1$.}
\label{EC1}
\end{figure}

\begin{figure}[H]
\centering
\includegraphics[width=3in]{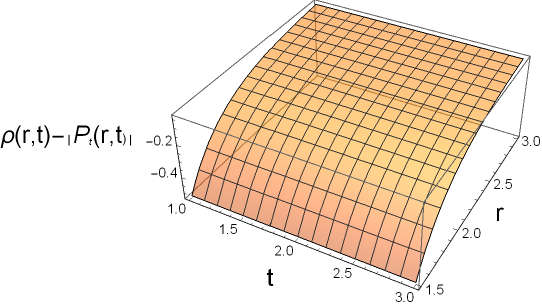}
\caption{The variation of DECs ($\rho(r,t)-\mid P_{t}(r,t)\mid$) for $R_0 = \alpha = 1$ and $\Lambda = 10^{-35}$ and the throat radius $r_0 = 1$.}
\label{EC2}
\end{figure}
Now, Let us check whether the matter is exotic or not by calculating some energy conditions namely
\
\begin{equation}
\rm WEC:\hspace {1.7 cm}\rho \geq 0 \hspace{0.5 cm} \rho+P_{r} \geq 0.
\label{46}
\end{equation}
\begin{equation}
\rm NEC:\hspace {1.1 cm}\rho+P_{r} \geq 0 \hspace{0.5 cm} \rho+P_{t} \geq 0.
\label{46}
\end{equation}
\begin{equation}
\rm DEC:\hspace {0.7 cm}\rho-\mid P_{r}\mid \geq 0 \hspace{0.5 cm} \rho-\mid P_{t}\mid \geq 0.
\label{46}
\end{equation}

In Figs.\ref{EC}, \ref{EC1} and \ref{EC2} we have plotted variation of the energy density $\rho(r,t)$, $\rho(r,t)+P_{r}(r,t)$ and $\rho(r,t)-\mid P_{t}(r,t)\mid$ for $r_0 = 1$. As can be seen from equations (\ref{43})--(\ref{45}) for $c_1 = \frac{1}{3}$ the WEC, NEC and DEC is violated throughout the space-time and so the matter is exotic for this case.

As we mentioned before, a fundamental ingredient of static traversable wormhole solutions in GR is the NEC violation.  However, for time-dependent wormhole solutions in GR the NEC and the WEC violations can be avoided for a specific interval of time and in certain regions at the throat \cite{Bordbar:2010zz, Heydari-Fard:2017oee, ev1, ev2, ev3}. Nevertheless, in some alternative gravity theories such as $f(R)$ gravity, Einstein-Gauss-Bonnet theory, Lovelock and Rastall gravity the energy conditions can be satisfied depending on the parameters of theory and thus the wormhole geometries can be constructed without any form of exotic matter. In these alternative gravity theories, similar to GR, the time-dependent spherically symmetric wormhole solutions have been extensively studied in the literature. For time-dependent wormhole solutions in $f(R)$ gravity theory the energy conditions are satisfied for the specific values of the model parameters \cite{fr}. But, this is not the case for time-dependent wormhole solutions analysed in this work.

In conformal Weyl gravity as a fourth-order gravitational theory both the static and time-dependent wormhole solutions differ from their counterparts in GR. For the static wormhole solution in Weyl gravity for example in the simple case of $b(r)=r_0$, in contrast to GR the radial pressure is positive at the throat and the energy density is negative, while similar to GR the NEC is violated throughout the space-time \cite{F. S. N. Lobo}. However, for the case of the time-dependent wormhole geometry studied here, we have proved hat the NEC and WEC is violated, as shown in the analysis above for the specific case $c_1 = \frac{1}{3}$. Finally, we mention that the restriction for choosing the constants results from the mathematical/technical reason not the physical one.

\section{Conclusion\label{sec4}}
There are two methods for formulating wormhole solutions. One method involves joining two asymptotically flat space-times via boundary conditions, while the other method involves smoothly merging the wormhole metrics with a cosmological background. In this paper, we employ the latter method and present a spherically inhomogeneous structure that smoothly joins with a cosmological background within the context of conformal Weyl gravity. Our ansatz metric belongs to the category of the Szekeres-Szafron metric, with two unknown functions, $a(r)$ and $R(t)$. Based on reasonable constraints on the energy-momentum tensor of an anisotropic space-time, we obtain the Weyl equations. These equations, together with resulting equations from the Bianchi and trace identities, i.e., $-\rho+P_r+2P_t=0$ make a set of equations, which have no exact solution in the general case. Considering two special cases, isotropic and anisotropic fluid, leads us to categories of equations based on the amount of $c_1$ as an integration constant. We obtain the de Sitter space-time as an exact solution which corresponds to $c_1=0$, $\rho=P_r=P_t=0$. Another exact solution has been obtained for special case $P_r = -P_t$, which is correspond to time-dependent wormhole for $c_1 = -\frac{1}{3}$, which can be supported by exotic matter which at large $r$ match two hyperbolic FRW universe.

\section*{Acknowledgements}
Malihe Heydari-Frad thanks the editor and also the anonymous referee for constructive suggestions that have improved the content and presentation of this paper.

\setcounter{equation}{0}
\renewcommand\theequation{A.\arabic{equation}}
\section*{Appendix A}

In this section we obtain the solution of equation (\ref{34}) for $\omega = -1$ by using MAPLE software as follows
\begin{eqnarray}
a(r) &=& -1 + \frac{1}{r^2}\times RootOf[1+2 c_3 r^2\nonumber\\
&+&2\int^{Z_{-}}\frac{df}{\sqrt{c_2f^{\frac{7}{2}}-12c_1f^4+4f^4}}r^2],
\label{Ap1}
\end{eqnarray}
which $c_2$ and $c_3$ are integration constants and function RootOf is a placeholder for representing all the roots of an equation in one variable. As one can see from equation (\ref{Ap1}) for $c_1= \frac{1}{3}$ the integrate can be solved easily and we fine the following exact solution
\begin{eqnarray}
c_2-\frac{c_3}{r^2}-\frac{8}{3}\frac{1}{
r^{3/2}(1+a(r))^{3/4}
}=0,
\label{Ap2}
\end{eqnarray}
or
\begin{equation}
a(r)=-1+\frac{1}{(\frac{3}{8})^{\frac{4}{3}}\left(c_2 r^{\frac{3}{2}}-c_3 r^{-\frac{1}{2}}\right)^{\frac{4}{3}}}.
\label{Ap2}
\end{equation}
Also, for the case of $c_2=0$, the equation (\ref{Ap1}) leads to the following solution
\begin{eqnarray}
a(r) = -1+\frac{1}{1-c_3 r^2+c_4},
\label{Ap3}
\end{eqnarray}
where $c_4=\sqrt{3 c_1+1}-1$. The above solution can not describe the wormhole solution since it does not satisfy the wormhole conditions and is not physically suitable. We do not discuss this solution in this paper.

\end{multicols}
\end{document}